\documentclass[a4paper,
               keeplastbox,   
               ]{jacow}
\usepackage{pdfpages,multirow,ragged2e} %
\makeatletter%
	\ifboolexpr{bool{xetex}}
	 {\renewcommand{\Gin@extensions}{.pdf,%
	                    .png,.jpg,.bmp,.pict,.tif,.psd,.mac,.sga,.tga,.gif,%
	                    .eps,.ps,%
	                    }}{}
\makeatother

\ifboolexpr{bool{xetex} or bool{luatex}} 
 {}                                      
 {\usepackage[utf8]{inputenc}}           

\usepackage[USenglish]{babel}

\ifboolexpr{bool{jacowbiblatex}}%
 {%
  \addbibresource{jacow-test.bib}
  \addbibresource{biblatex-examples.bib}
 }{}
\begin{document}

\title{Recent Results from the Study of Emittance Evolution in MICE}

\author{V. Blackmore\thanks{v.blackmore@imperial.ac.uk}, Imperial College London, London, UK \\
		on behalf of the MICE Collaboration}
	
\maketitle

\begin{abstract}
   The Muon Ionization Cooling Experiment (MICE) has measured the evolution of emittance due to ionization energy loss. Muons were focused onto an absorber using a large aperture solenoid. Lithium-hydride and liquid hydrogen-absorbers have been studied. Diagnostic devices were placed upstream and downstream of the focus, enabling the phase-space coordinates of individual muons to be reconstructed. By observing the properties of ensembles of muons, the change in beam emittance was measured. Data taken during 2016 and 2017 are currently under study to evaluate the change in emittance due to the absorber for muon beams with various initial emittance, momenta, and settings of the magnetic lattice. The current status and the most recent results of these analyses will be presented.
\end{abstract}

\section{Introduction}

Low emittance stored muon beams have been proposed as the source of neutrinos at a neutrino factory~\cite{NF1,NF2} or as the means to deliver multi-TeV lepton-antilepton collisions at a muon collider~\cite{MC1}. The muons at such a facility originate from the decay of pions and occupy a large volume in phase space (emittance). This must be reduced (cooled) to fit within the acceptance of a storage ring or accelerating structure.  The short muon lifetime prohibits existing cooling techniques such as synchrotron or stochastic cooling. Ionisation cooling is the only practical method of cooling muon beams.

Ionisation cooling occurs when a muon beam passes through an absorber material and ionises atomic electrons, losing both transverse and longitudinal momentum. This reduces the volume in phase space occupied by the beam, and thus reduces the beam emittance. Concurrently, multiple Coulomb scattering within the absorber increases the angular divergence of the beam, increasing the emittance. The interplay of emittance reduction, due to energy loss by ionisation, and emittance growth, through multiple Coulomb scattering, results in a change in normalised RMS emittance, $\varepsilon_{\perp}$, over a distance, $dz$, given by~\cite{coolingEq},
\begin{equation}
    \frac{d\varepsilon_{\perp}}{dz} \approx -\frac{\varepsilon_{\perp}}{\beta^{2}E_{\mu}} \left\langle \frac{dE}{dz} \right\rangle + \frac{\beta_{\perp} (\mathrm{13.6\,MeV/}c)^{2}}{2 \beta^{3} E_{\mu} m_{\mu} X_{0}}, 
    \label{eq:coolingEq}
\end{equation}
where $\beta_{\perp}$ is the transverse optical Twiss function, $\beta c$, $E_{\mu}$, and $m_{\mu}$ are a particle's velocity, energy, and mass respectively, and $X_{0}$ is the radiation length of the absorber material. Maximum emittance reduction is achieved by using absorbers such as liquid hydrogen or lithium hydride.

\section{The Muon Ionisation Cooling Experiment}

\begin{figure*}
\vspace{-12em}
    \centering
    \includegraphics*[width=0.9\textwidth]{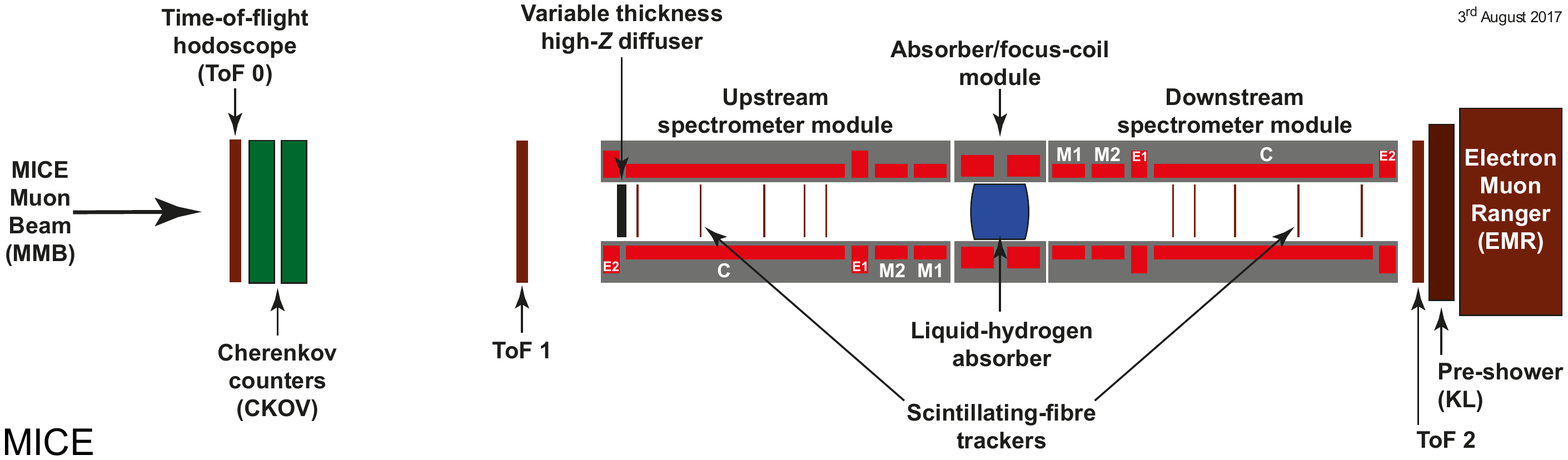}
    \vspace{-8em}
    \caption{The instrumentation of MICE. Single muons cross two Time-of-Flight detectors and Cherenkov detectors, then enter the solenoidal field of the MICE magnetic lattice. 
The muon's position and momentum is measured in the upstream solenoid module using a scintillating fibre tracker. The muon then crosses an absorber, losing momentum, and is re-measured in the downstream solenoid module with a scintillating fibre tracker. The muon is instrumented on its exit with Time-of-Flight, Kloe-lite and Electron-Muon Ranger detectors.
    }
    \label{fig:mice}
\end{figure*}

The Muon Ionisation Cooling Experiment (MICE)~\cite{miceExp} consists of a transfer line that transports particles from the ISIS synchrotron at the Rutherford Appleton Laboratory, UK, to a segment of a cooling lattice. The beam is transported through PID detectors using conventional quadrupole magnets, then by large-aperture superconducting solenoids operated at up to 3\,T. At the centre of the cooling lattice is an absorber material. The beam is measured before and after the absorber in the constant 3\,T field of the up- and downstream spectrometer solenoid modules using scintillating fibre trackers. A schematic drawing of MICE is shown in Figure~\ref{fig:mice}.

MICE has measured the passage of individual muons through an empty channel, 21 litres of liquid hydrogen contained within an aluminium vessel, and a 65\,mm solid lithium hydride disc. These measurements were made with the same magnetic optical configuration to evaluate phase space evolution across a cooling cell and compare the cooling efficiency of liquid hydrogen and lithium hydride. Individual particle trajectories are measured as they cross the experiment and are assembled offline into analysable particle ensembles. These ensembles consist of particles with: 1) an upstream Time-of-Flight consistent with a muon, 2) a measured total momentum, $p$, in the range 135--145\,MeV/$c$, 3) a single, good quality track in the upstream tracking detector, and 4) a measured momentum in the upstream tracking detector consistent with the Time-of-Flight.
The current status of the analysis of the measured ensembles is presented here.

\section{Emittance evolution}

\begin{figure}
    \centering
    \includegraphics[width=0.49\columnwidth]{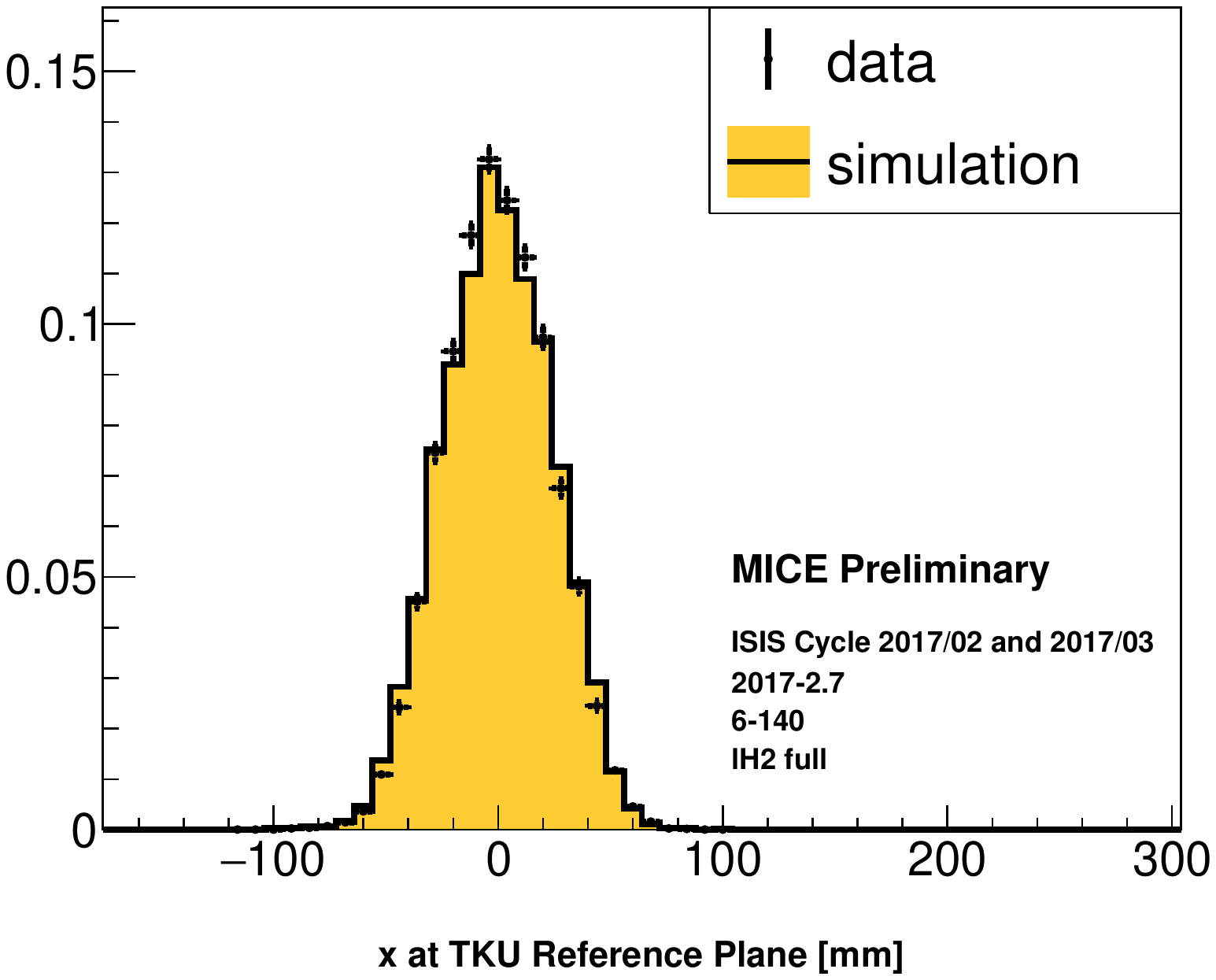} ~\includegraphics[width=0.49\columnwidth]{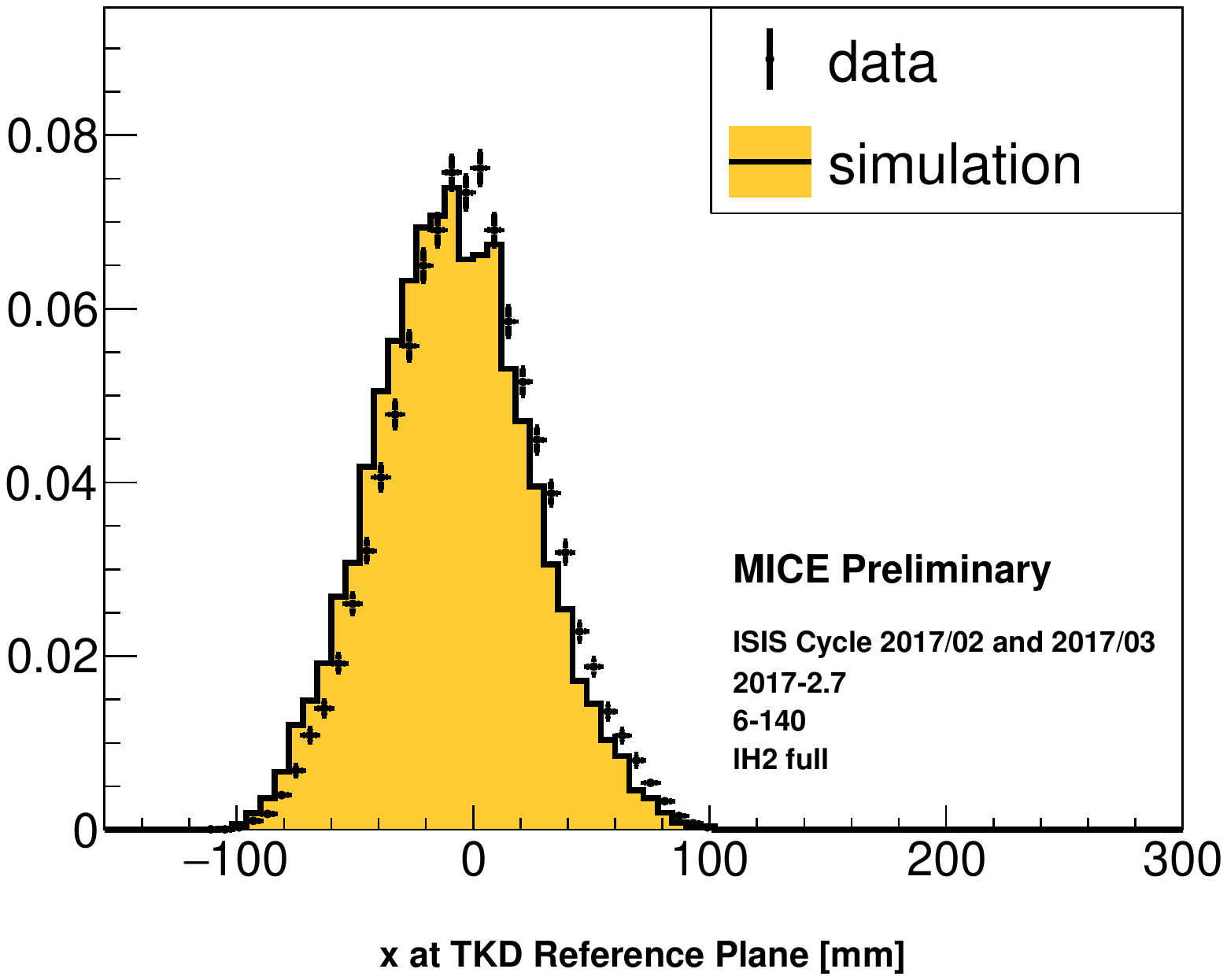} \\
        \includegraphics[width=0.49\columnwidth]{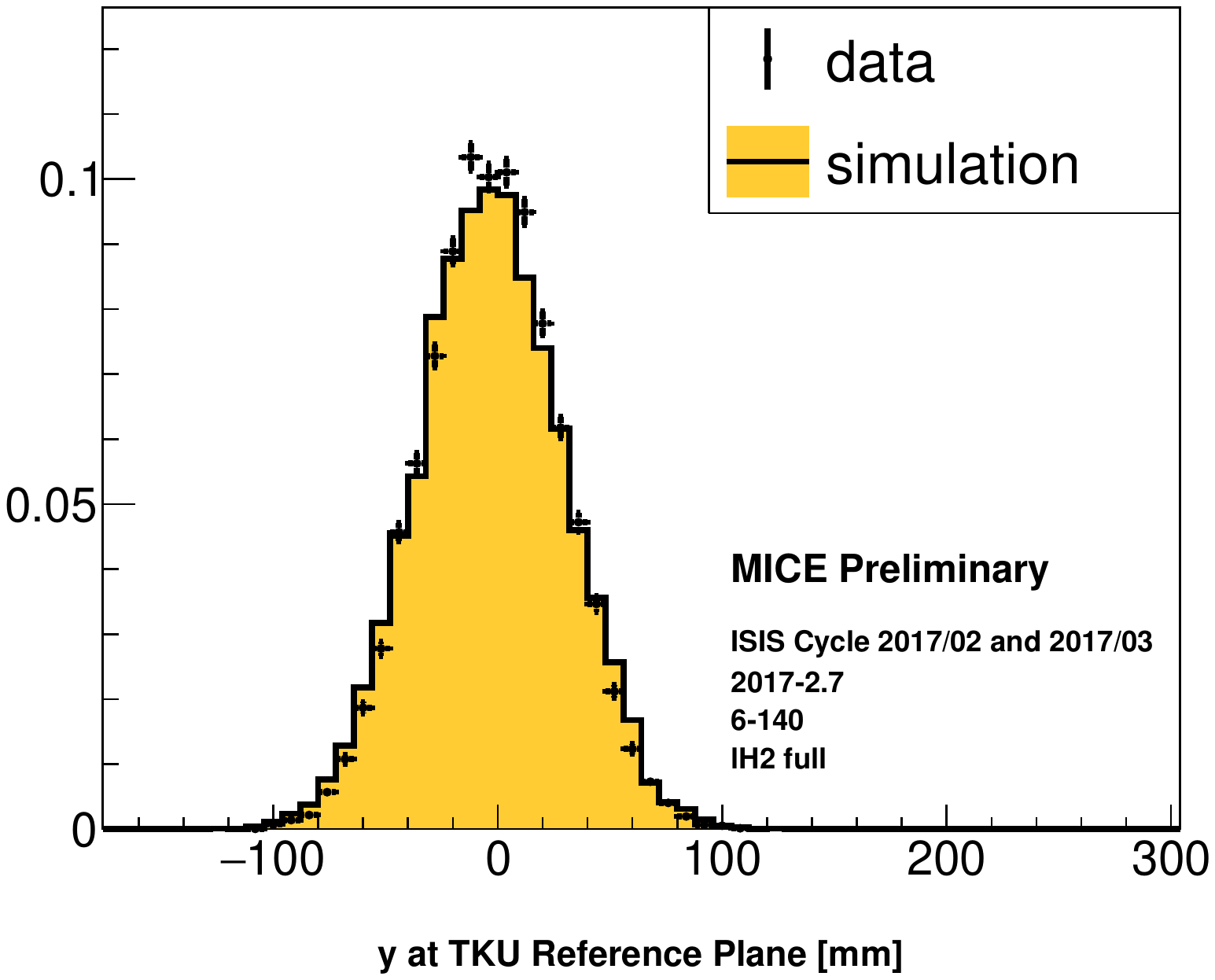}~\includegraphics[width=0.49\columnwidth]{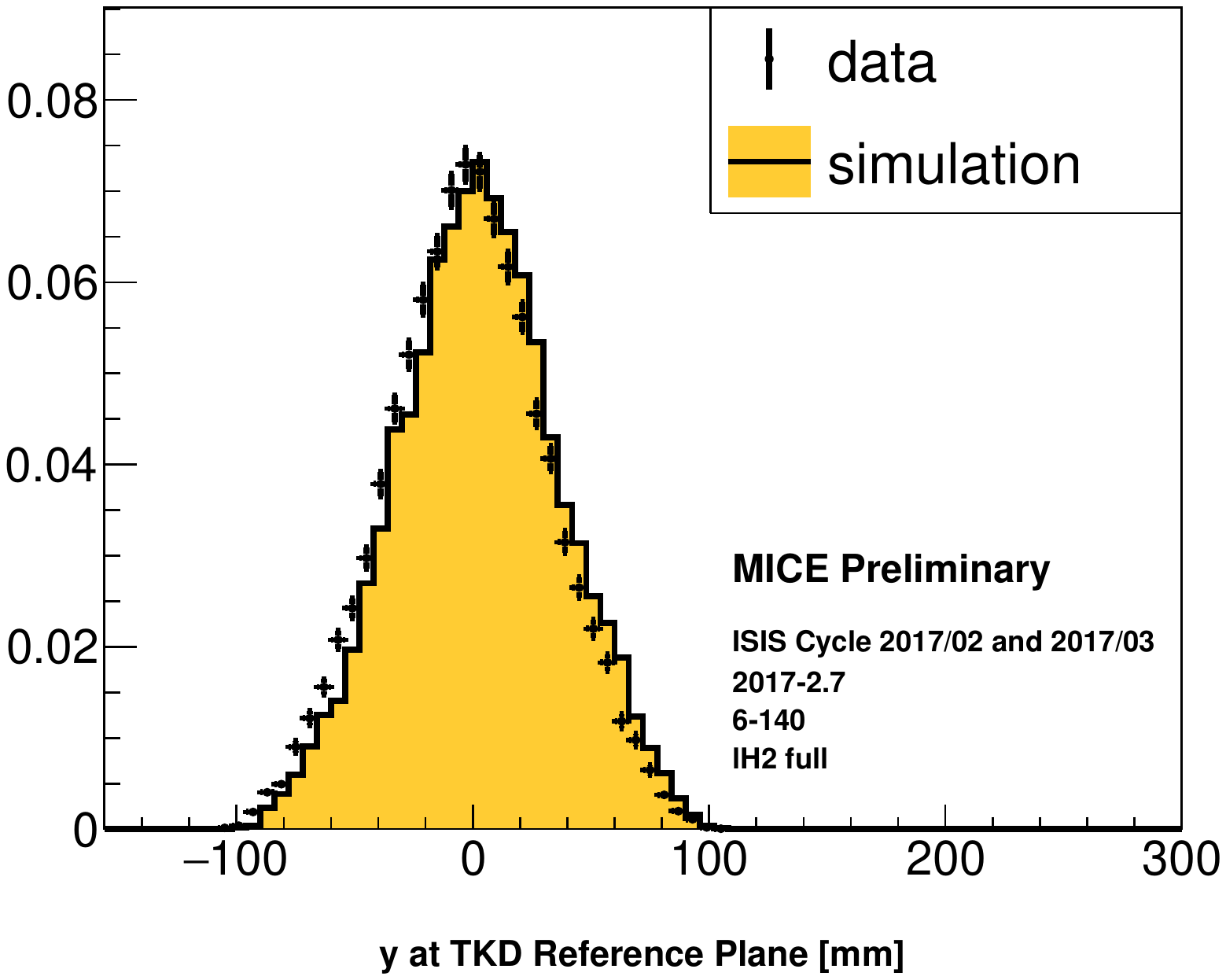} \\
            \includegraphics[width=0.49\columnwidth]{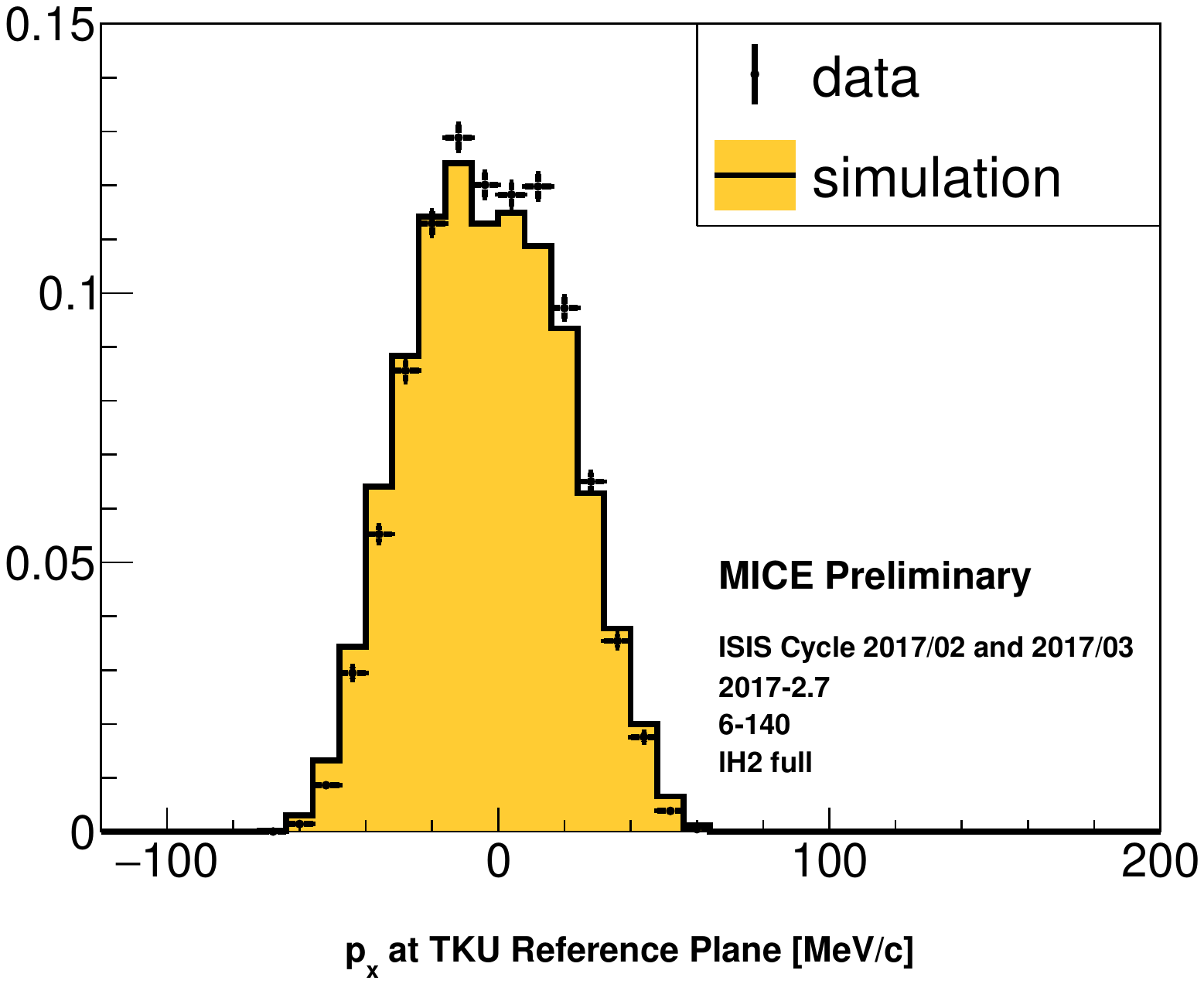}~\includegraphics[width=0.49\columnwidth]{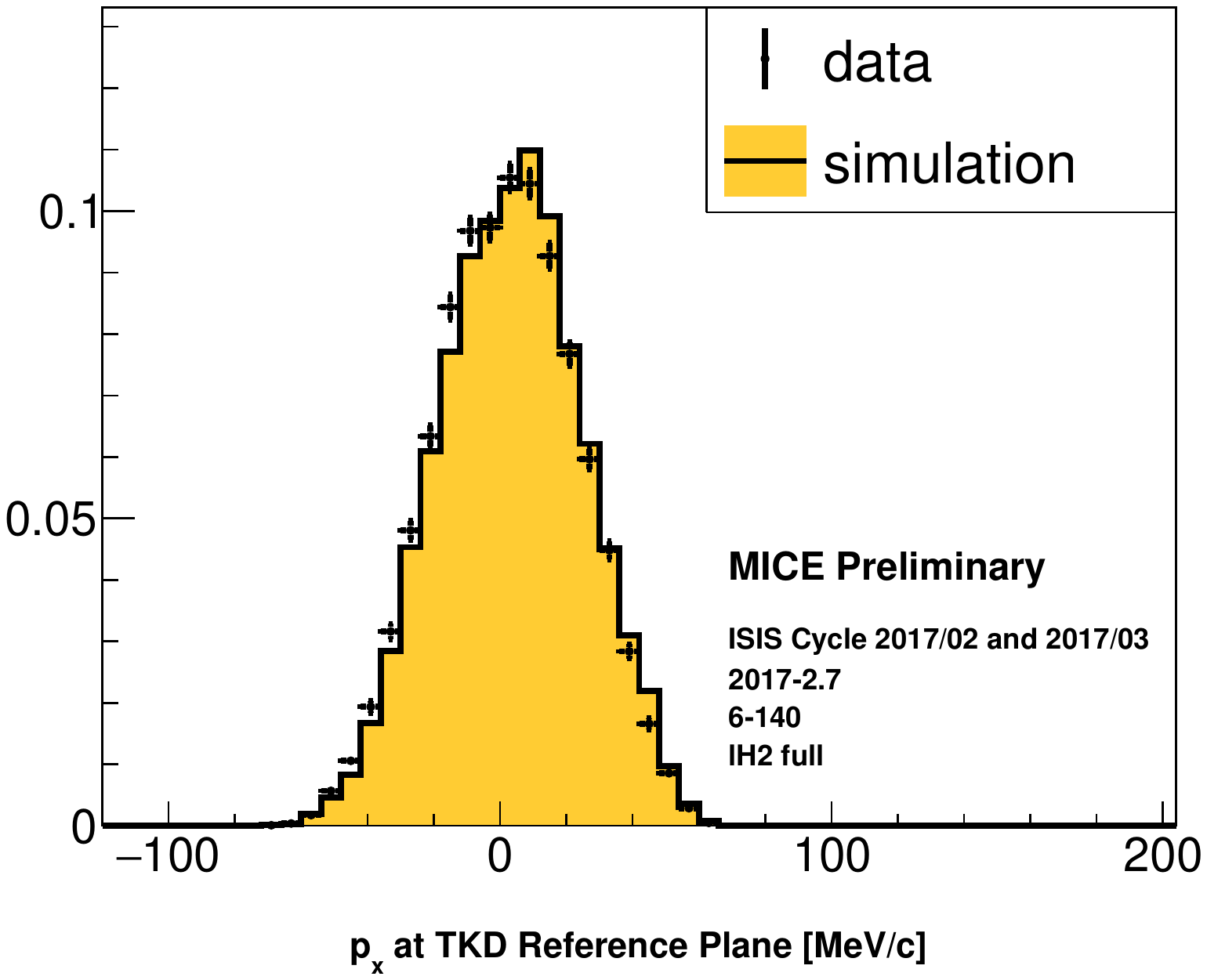} \\
                \includegraphics[width=0.49\columnwidth]{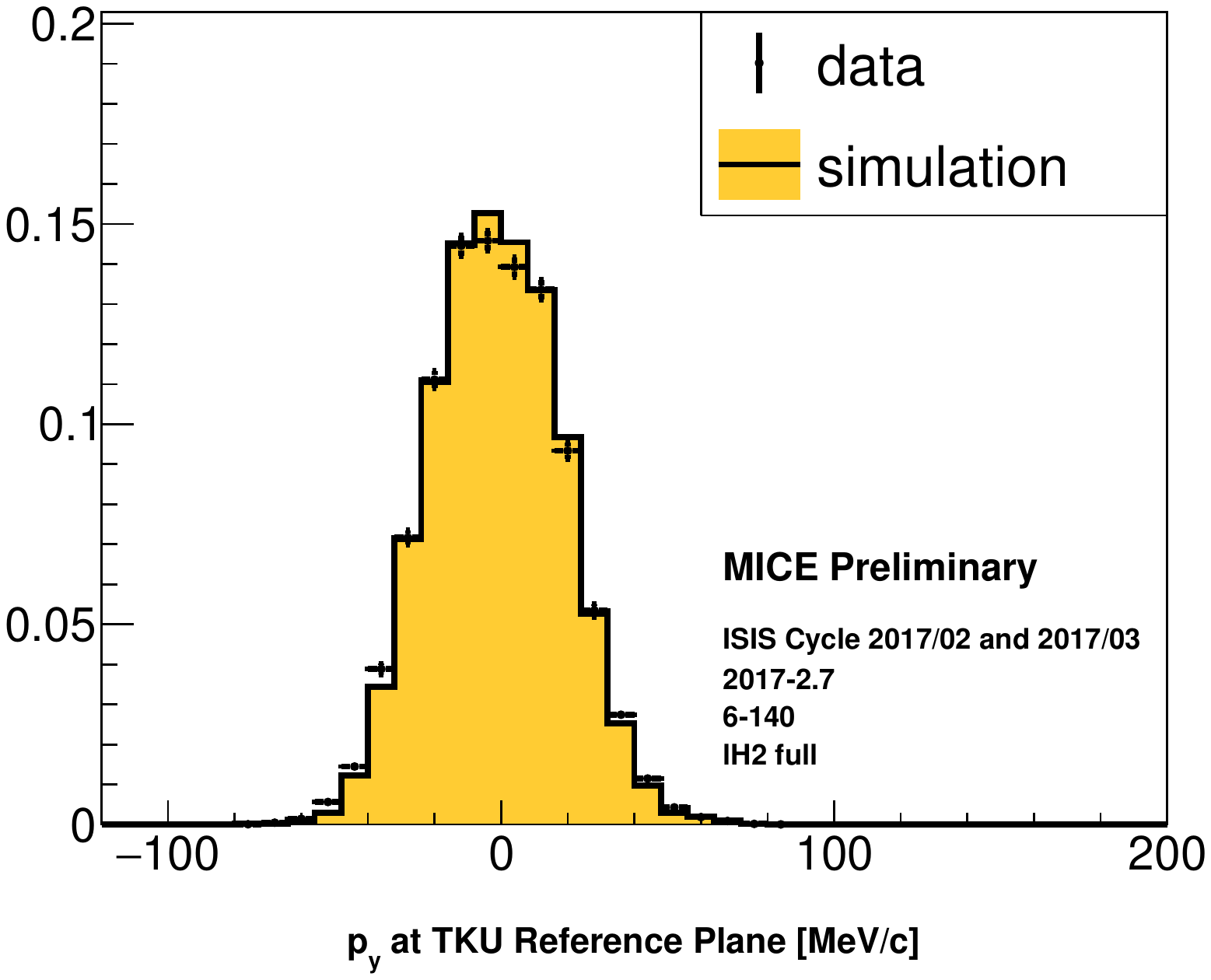}~\includegraphics[width=0.49\columnwidth]{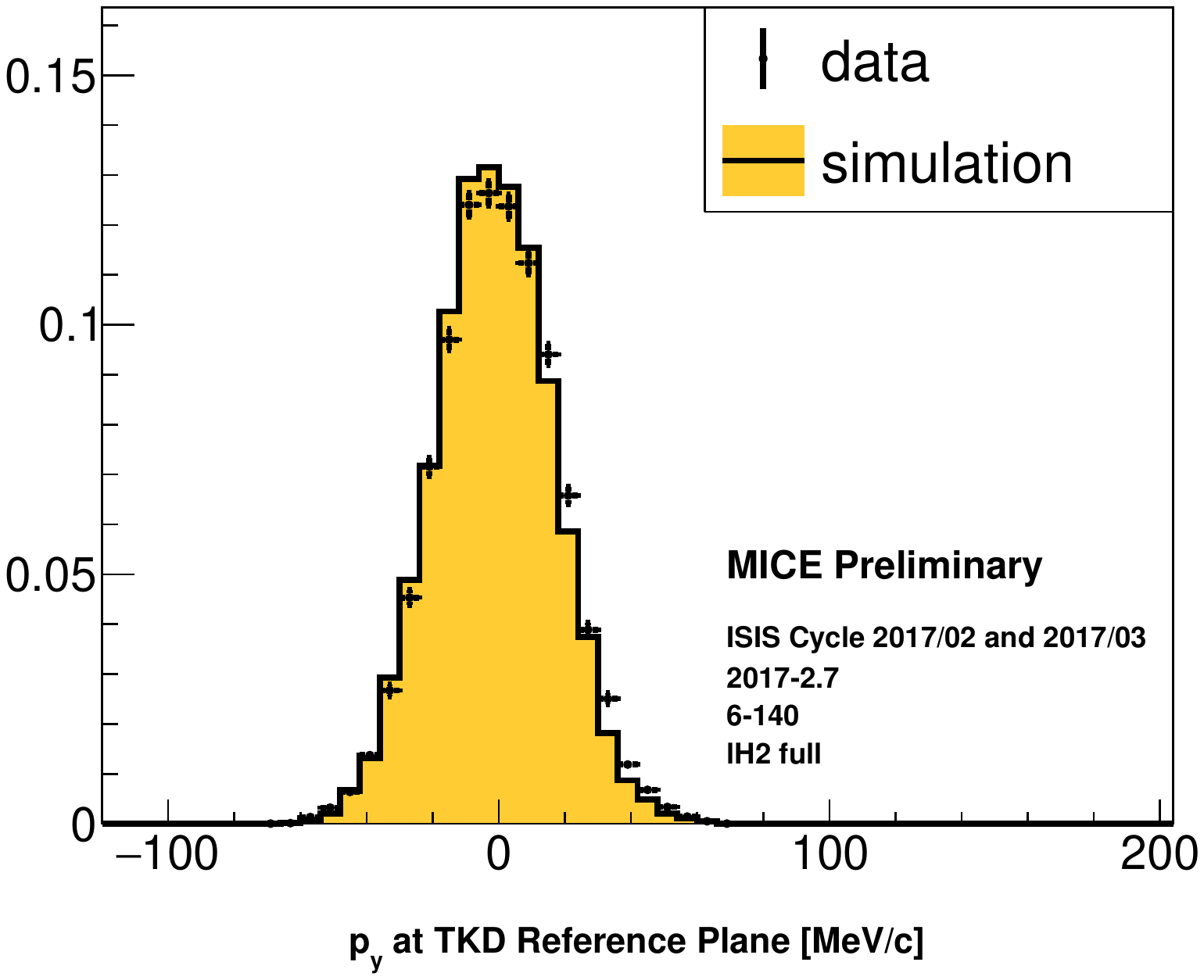} \\
    \caption{Comparison of reconstructed data (black, circles) and reconstructed Monte Carlo simulations (filled histogram) of muon beam distributions (left) upstream and (right) downstream of a liquid hydrogen absorber. Distributions are, from top to bottom, $x, y, p_{x}, p_{y}$.  These are the components required to calculate the transverse amplitude of a particle.}
    \label{fig:beam-distrbution}
\end{figure}

The measured components of transverse phase space, $x, y, p_{x}, p_{y}$ are compared to those of a Monte Carlo simulation, that includes the response of instrumentation, in Figure~\ref{fig:beam-distrbution}.  The reconstructed data (black, circles) and the reconstructed simulation (filled histogram) show good agreement. Distributions of $x, y, p_{x}, p_{y}$ are shown (left) up- and (right) downstream of the liquid hydrogen absorber. 

\begin{figure*}
\vspace{-5em}
    \centering
    \includegraphics*[width=0.9\textwidth]{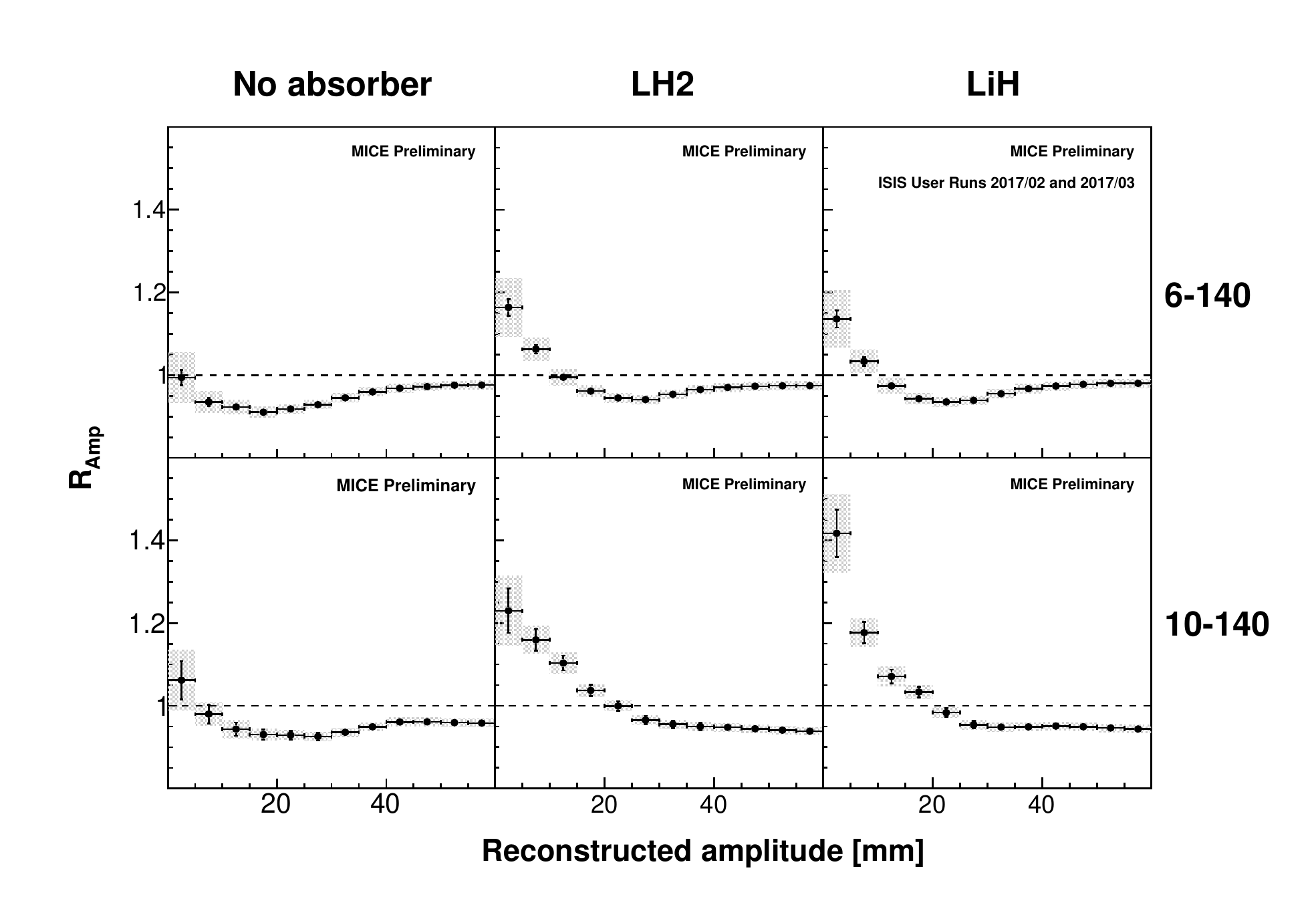}

    \caption{
    The ratio, $R_{\mathrm{Amp}}$, of the downstream and upstream samples for (left) an empty channel, (middle) liquid hydrogen, and (right) lithium hydride. The ratios are shown for two input beam line settings, with a nominal momentum of $p=140$\,MeV/$c$ and nominal emittance of 6 (top), and 10\,mm (bottom). Cooling is seen where $R_{\mathrm{Amp}} > 1$. Statistical uncertainties are given as bars, systematic uncertainties (grey) are under study.
    }
    \label{fig:cooling}
\end{figure*}

The ensemble's covariance matrix, $\Sigma_{\perp}$, is calculated from the measured transverse quantities as,
\begin{equation}
  \Sigma_{\perp} = \left(
    \begin{matrix}
      \sigma_{xx} & \sigma_{xp_x} & \sigma_{xy} & \sigma_{xp_y} \\
      \sigma_{xp_x} & \sigma_{p_{x}p_{x}} & \sigma_{yp_x} & \sigma_{p_{x}p_{y}} \\
      \sigma_{xy}   & \sigma_{yp_x} & \sigma_{yy} & \sigma_{yp_y} \\
      \sigma_{xp_y} & \sigma_{p_{x}p_{y}} & \sigma_{yp_y} & \sigma_{p_yp_y}
    \end{matrix}
  \right),
  \label{Eq:Var4D}  
\end{equation}
and the transverse RMS emittance is,
\begin{equation}
  \varepsilon_{N} = \frac{1}{m_{\mu}}\sqrt[4]{\det  \Sigma_{\perp}}.
  \label{Eq:Norm4D}
\end{equation}
The transverse amplitude, $A_{\perp}$, of a particle with respect to the ensemble is,
\begin{equation}
    A_{\perp} = \varepsilon_{N}(\mathbf{v} - \mathbf{\mu})^{T}\Sigma_{\perp}^{-1}(\mathbf{v} - \mathbf{\mu})
    \label{Eq:amp}
\end{equation}
where $\mathbf{v}$ is the phase space vector of the particle, $\mathbf{v} = (x, p_{x}, y, p_{y})$, and $\mathbf{\mu} = (\langle x \rangle, \langle p_{x} \rangle , \langle y \rangle , \langle p_{y} \rangle )$ is the beam centroid.
An ensemble of muons with large emittance, consists of a large spread of particle amplitudes, whereas an ensemble with small emittance has a more particles occupying lower amplitudes.

Data were taken with the same configuration of solenoidal fields and input beam, i.e $\beta_{\perp}\approx 660$\,mm at the absorber. Measurements were made of the particles that crossed the experiment up- and downstream of the absorbers. As the magnetic field was the same in each data set  
the empty channel measurements enable observation of optical abberrations and will be used to cross-check for other systematic effects.
Additionally in the case where Liouville's theorem holds true, such as for an empty channel, emittance should be conserved between the up- and downstream reference surfaces of the tracking detectors.

The amplitude of particles measured up- and downstream in each case (empty channel, liquid hydrogen, lithium hydride) were calculated according to Eq.~\ref{Eq:amp}. These amplitudes were binned in 5\,mm increments, from $0 \leq A_\perp \leq 60$\,mm. 
In order to prevent sensitivity to tail effects, the covariance matrix is recalculated excluding all higher-amplitude particles for each amplitude bin. 
The number of muons within each amplitude bin in the upstream and downstream samples was calculated. An increase in the number of low-amplitude particles downstream implies an increase in density of the beam core, or a lower emittance.

The cumulative transverse amplitude was calculated for each use case (empty channel, liquid hydrogen, lithium hydride) and the ratio of the number of muons in the downstream to upstream samples was taken as 
\begin{equation}
R_{\mathrm{Amp}} = \frac{N_{\mathrm{downstream}}}{N_{\mathrm{upstream}}},
\end{equation}
where $N$ is the number of muons with amplitude equal to or less the amplitude under consideration. Figure~\ref{fig:cooling} shows the variation of this ratio with increasing amplitude, encompassing increasing amounts of the ensemble, for the empty channel, liquid hydrogen, and lithium hydride absorbers. Two different input beams are also shown, with a nominal $\varepsilon_{\perp,\mathrm{upstream}} \approx 6$ and $10$\,mm. If the ensemble is cooled, the ratio $R_{\mathrm{Amp}} > 1$, as is seen in both liquid hydrogen and lithium hydride absorbers. No cooling is seen in the empty channel ensemble.


\section{Conclusion}

MICE has measured a reduction in the transverse amplitude of muons crossing liquid hydrogen and lithium hydride absorbers. Particles observed with large transverse amplitude upstream of the absorber material migrate to lower amplitude areas after crossing the absorber. No emittance reduction is seen with the same optical field configuration but without an absorber in place.  The Monte Carlo simulation is representative of the transverse distributions of the ensemble.  Further analysis is in progress, including a robust estimate of systematic uncertainties due to field non-uniformity and other sources. Additional analyses are underway to study the equilibrium emittance of each absorber material, and the dependence of Eq.~\ref{eq:coolingEq} on momentum and the value of $\beta_{\perp}$ at the absorber.

\section*{Acknowledgements}

The work described here was made possible by grants from Department of Energy and National Science Foundation (USA), the Instituto Nazionale di Fisica Nucleare (Italy), the Science and Technology Facilities Council (UK), the European Community under the European Commission Framework Programme 7 (AIDA project, grant agreement no. 262025, TIARA project, grant agreement no. 261905, and EuCARD), the Japan Society for the Promotion of Science and the Swiss National Science Foundation, in the framework of the SCOPES programme. 
We gratefully acknowledge all sources of support. We are grateful to the support given to us by the staff of the STFC Rutherford Appleton and Daresbury Laboratories. We acknowledge the use of Grid computing resources deployed and operated by GridPP in the UK, \url{http://www.gridpp.ac.uk/}.

%
%
\ifboolexpr{bool{jacowbiblatex}}%
	{\printbibliography}%
	{%
	

} 

\end{document}